\documentclass[aps,twocolumn,prl,showpacs,superscriptaddress,groupedaddress]{revtex4}

\usepackage{dcolumn}
\usepackage{bm}
\usepackage{color}

\def\ltape{\hbox{\ $<$\hskip -8pt\raise -4pt\hbox{$\sim$}\ }}
\def\gtape{\hbox{\ $>$\hskip -8pt\raise -4pt\hbox{$\sim$}\ }}

   \ifx\pdfoutput\undefined
   \usepackage[dvips]{graphicx}
   \else
   \usepackage[pdftex]{graphicx}
   \fi
   \usepackage{epstopdf}
   \DeclareGraphicsExtensions{.pdf,.gif,.jpg,.eps,.png}
  
   \usepackage{amsmath}
  \usepackage{amssymb}

\begin{document}


\title{Scaling of magnetic reconnection with a limited x-line extent}

\author{Kai~Huang}
\affiliation{University of Science and Technology of China, Hefei, China}
\affiliation{Dartmouth College, Hanover, NH 03750}
\author{Yi-Hsin~Liu}
\affiliation{Dartmouth College, Hanover, NH 03750}
\author{Quanming~Lu}
\affiliation{University of Science and Technology of China, Hefei, China}
\author{Michael~Hesse}
\affiliation{University of Bergen, Bergen, Norway}

\date{\today}

\begin{abstract}
Contrary to all the 2D models, where the reconnection x-line extent is infinitely long, we study magnetic reconnection in the opposite limit. The scaling of the average reconnection rate and outflow speed are modeled as a function of the x-line extent. An internal x-line asymmetry along the current direction develops because of the flux transport by electrons beneath the ion kinetic scale, and it plays an important role in suppressing reconnection in the short x-line limit; the average reconnection rate drops because of the limited active region, and the outflow speed reduction is associated with the reduction of the $J \times B$ force, that is caused by the phase shift between the J and B profiles, also as a consequence of this flux transport.   
\end{abstract}

\pacs{52.27.Ny, 52.35.Vd, 98.54.Cm, 98.70.Rz}

\maketitle

{\it Introduction--}
Magnetic reconnection is a process that re-organizes the global magnetic geometry and explosively releases magnetic energy in planetary magnetospheres \cite{dungey61a, angelopoulos08a}, solar wind \cite{gosling05a, phan06a}, solar flares \cite{giovanelli1946theory, Masuda94}, and potentially in astrophysical systems \cite{zweibel09a, uzdensky2011magnetic}. The basic idea of magnetic reconnection can be captured in ``two-dimensional (2D)'' picture, that involves the break and rejoining of magnetic field lines on a 2D plane. However, two-dimensional models and simulations of magnetic reconnection impose an unnecessary condition that enforces quantities to be translational invariant in the out-of-plane direction. Consequentially, reconnection x-line is {\it infinitely} long. This same situation also carries over to most three-dimensional (3D) simulations that have a uniform initial current sheet and periodic boundary condition in the x-line direction. While results from a 2D model may be sufficient in some applications [e.g.,\cite{TKMNakamura18a}], some incidences suggest that reconnection is operating in the opposite limit, and its property is far less understood.

Many observations and simulations indicate that the thin current sheet and reconnection x-line can be spatially confined, for instance, in the magnetotails of planets (like Mercury \cite{slavin2009messenger, chen2019studying, dong2019novel}), moons (like Ganymede \cite{dorelli2015role}) and comets \cite{russell1986near}, where the spatial scale in the cross-tail direction is short compared to plasma kinetic scales. Even in planets with larger magnetotails, like Earth, the thin current sheet necessary for magnetic reconnection onset can also be spatially limited; the spatial scale may be determined by the size of external drivers \cite{nishimura2016localized}, the wavelength of internal instabilities \cite{pritchett2011plasma, pritchett14a}, or the inhomogeneity of the neutral sheet caused by other global effects \cite{baker82a}. The resulting spatially confined reconnection could be relevant to the generation of Dipolarizing Flux Bundles (DFBs) \cite{JLiu15b}, or Bursty Bulk Flows (BBFs) \cite{nakamura04a, nagai13a}, that have short cross-tail extents. Thus reconnection with a limited x-line extent is likely ubiquitous, and it is imperative to study its nature. In this letter, we quantitatively model the scaling of reconnection rate and the ion outflow speed as a function of the x-line extent. Notably, these two quantities are essential in defining the well-being of magnetic reconnection.

{\it Simulation setup--}
The setup of simulations is detailed in Liu et al., \cite{yhliu19a}, that is a modified Harris sheet with a thin current sheet of thickness $1d_i$ embedded between thick current sheets of thickness $8d_i$. Here $d_i$ is the ion inertial length based on the density of the sheet component $n_0$. Reconnection proceeds with a limited x-line extent that does not spread into the thick sheets. The length of this thin current sheet $L_{y,thin}$ (i.e., the x-line extent) is a free parameter we can control. $B_0 \hat{x}$ is the reconnecting component, Alfv\'en speed $V_A\equiv B_0/(4\pi m_i n_0)^{1/2}$ and ion gyro-frequency $\Omega_{ci} \equiv eB_0/m_ic$, where the ion to electron mass ratio $m_i/m_e=25$. Initial perturbation of strength $\delta B_z=0.05B_0$ is used to induce reconnection. Simulations are performed within boxes of size $L_x \times L_y \times L_z = 32d_i \times 64d_i \times 16d_i$ and $640 \times 1280 \times 320$ cells. The boundary conditions are periodic both in the x- and y-directions, while in the z-direction (normal to the current sheet) they are conducting for fields and reflecting for particles.

\begin{figure}
\includegraphics[width=8.5cm]{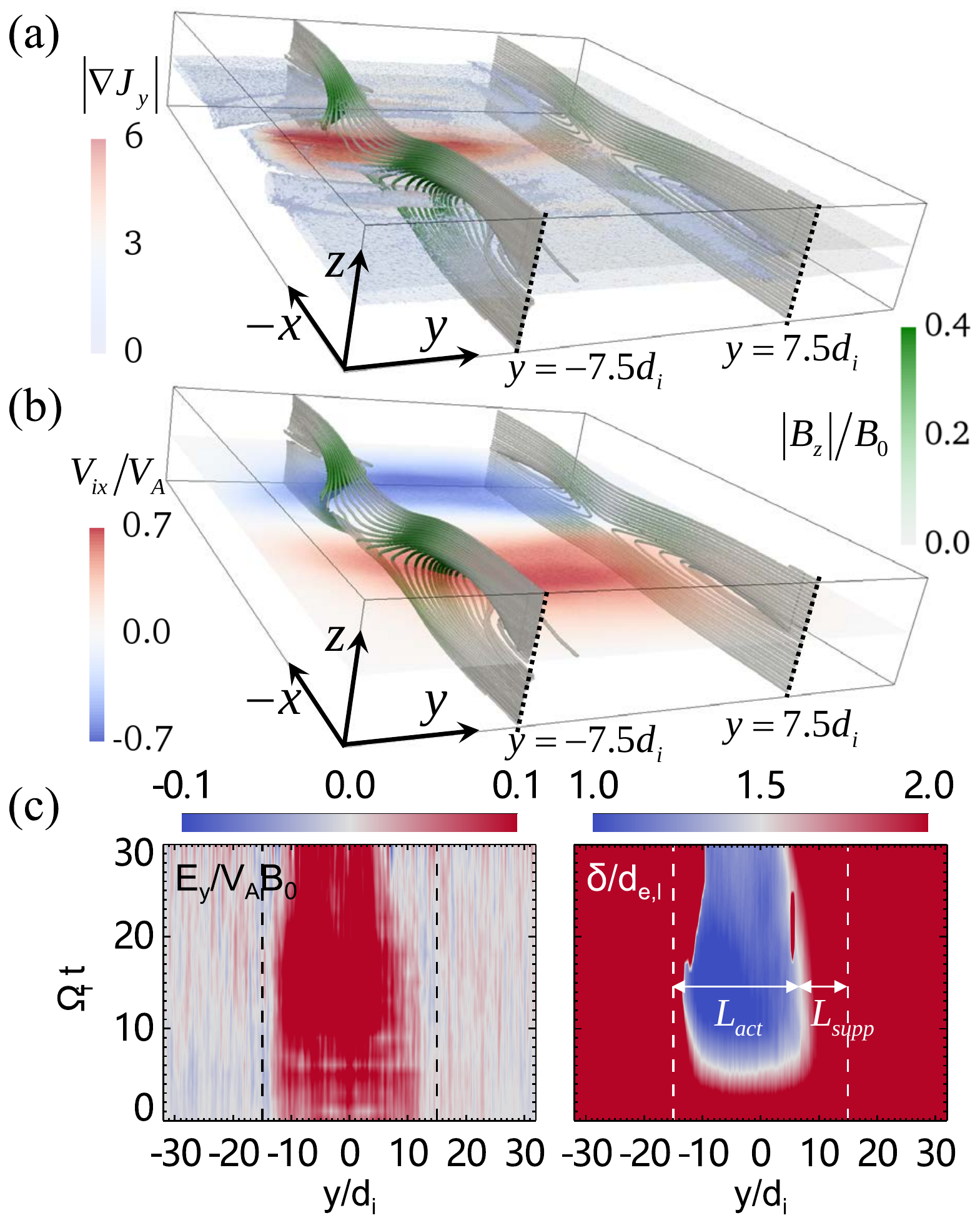} 
\caption {An overview of the $L_{y,thin}=30d_i$ case at time $15/\Omega_{ci}$. The 3D domain shown here is $[-16,16]\times [-15,15] \times [-3,3]d_i$. (a) The contour of the current density $J_y$ colored by its gradient $|\nabla J_y|$. (b) Ion outflow velocity $V_{ix}$ on the $z=0$ plane. Sample magnetic field lines at $y\simeq \pm7.5d_i$ are colored by the magnitude of $|B_z|$. (c) Time stacks of the electric field $E_y$ and the currents sheet half-thickness $\delta$ along the x-line of extent $L_{y,thin}$, that is bounded between the two vertical dashed lines.}
\label{FG1}
\end{figure}

{\it Overview of the morphology--}
Figure 1 displays features of magnetic reconnection with a limited x-line extent $\simeq 30d_i$ in the y-direction. Fig.~1(a) shows the contour of the current density $J_y$ that is colored by its gradient $|\nabla J_y|$. In the reddish region, the current sheet is thinner. Fig.~1(b) shows the ion outflow velocity $V_{ix}$ on the $z=0$ plane. It is clear that the active region with a thinner current sheet and large ion outflow velocity appears on the dawn side ($-y$ direction), while on the dusk side ($+y$ direction) the reconnection is suppressed, forming the ``internal x-line asymmetry''. 3D magnetic field lines across the active region at $y\simeq -7.5 d_i$ and the suppression region at $y\simeq 7.5 d_i$ are also plotted with the color showing the local strength of reconnected $B_z$. This internal x-line asymmetry develops because $B_z$ is transported to the dawn side by the electron diamagnetic and $E\times B$ drifts \cite{yhliu19a, yhliu16a}. In this paper, we will model how this flux transport affects the reconnection rate and outflow speed.

We quantify the active region of the x-line as the segment where the current sheet half-thickness is thinner than two local electron inertial length, $\delta < 2d_{e,l}$, that is essentially the electron diffusion region. Fig.~1(c) shows the time stacks of the reconnection electric field $E_y$ and the current sheet half-thickness $\delta$ along the x-line. We can see that the active region $\delta < 2d_{e,l}$ (in blue) coincides with the region with large $E_y$ (in red) that reaches a value $\simeq 0.1 V_A B_0$, the typical fast reconnection rate \cite{yhliu17a,TKMNakamura18a, genestreti18a}.
The thin current sheet consists of two distinct regions: the active region and the suppression region. Thus $L_{y,thin}=L_{act}+L_{supp}$ and $L_{supp}$ is determined to be $\sim \mathcal{O}(10d_i)$ \cite{yhliu19a}, as also seen in Fig.~1(c). 

\begin{figure}
\includegraphics[width=8.5cm]{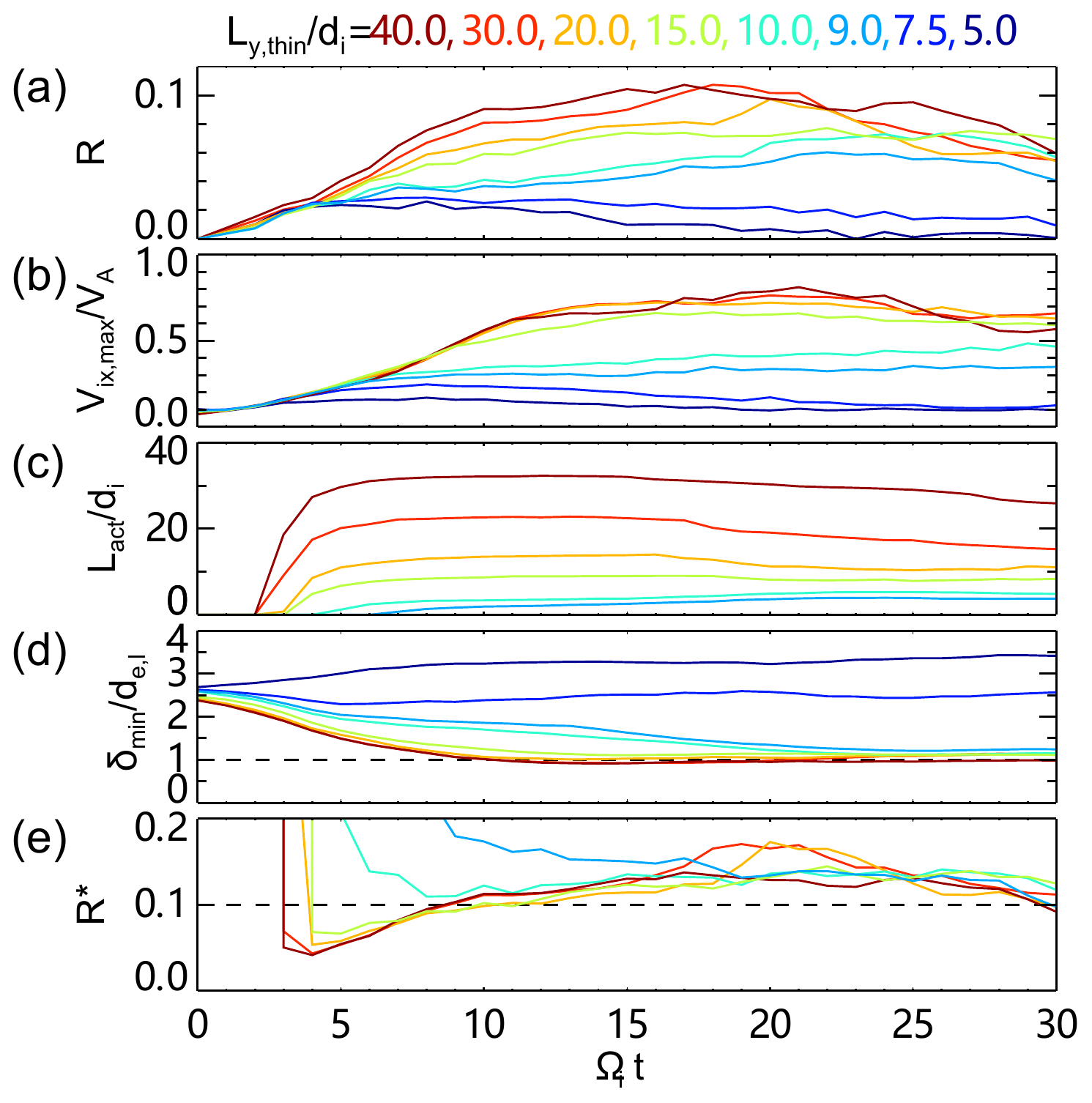} 
\caption {Time evolution of (a) the average reconnection rate $R$, (b) maximum ion outflow speed $V_{ix,max}$, (c) the active region extent $L_{act}$, (d) minimum half-thickness of the current sheet $\delta_{min}$, and (e) the reconnection rate normalized by the active region extent $R^*$. Cases with $L_{y,thin}/d_i=40$, 30, 20, 15, 10, 9, 7.5, and 5 are shown in different colors. $L_{y,thin}/d_i=7.5$ and $5$ cases are not shown in (c) and (e) because there is no active region.}
\label{FG2}
\end{figure}

{\it Scaling of the reconnection rate--}
The time evolution of important quantities in all the cases are given in Fig.~2. Panels (a) and (b) show the average reconnection rate $R$ and maximum ion outflow velocity $V_{ix,max}$, respectively. Here $R\equiv \partial_t \Psi/(V_AB_0L_{y,thin})$, where $\Psi=-\int_{-L_y}^{L_y}\int_{-L_x}^{X(y)}B_z(x,y,0)dxdy$ is the reconnected flux and $X(y)$ is the location of the primary x-line that maximizes $\Psi(X)=-\int_{-L_x}^{X}B_z(x,y,0)dx$ at a given y. It is evident that both reconnection rate and ion outflow velocity are strongly suppressed when $L_{y,thin}$ is shorter than $\simeq 10d_i$. Fig.~2(c) and (d) show the time evolution of the active region $L_{act}$ and the minimum half-thickness of the current sheet $\delta_{min}$, respectively. For the cases with $L_{y,thin}>10d_i$, a quasi-steady active region exists and $\delta_{min}$ reaches the local electron inertial length $d_{e,l}$, where the electron frozen-in condition is broken. For the two cases with $L_{y,thin}< L_{supp}\simeq 10d_i$, there is no active region; thus the current sheet does not thin much. It is clear that the current sheet thinning is important in facilitating fast reconnection, and the critical scale is $d_{e,l}$.

In Fig.~2(e), we renormalize the reconnection rate by the extent of the active region, $R^*\equiv \partial_t \Psi/(V_AB_0L_{act})$. Surprisingly, cases with an active region all appear to have a similar rate $R^*\simeq \mathcal{O}(0.1)$. In light of this observation, we can model the averaged reconnection rate as
\begin{equation}
R=\frac{L_{act}}{L_{y,thin}}R^*\simeq \left( 1-\frac{L_{supp}}{L_{y,thin}}\right) R^*
\label{rate}
\end{equation}
where $L_{supp}\simeq \mathcal{O}(10d_i)$. Note that Eq.(\ref{rate}) cannot explain the small but finite reconnection rate observed in $L_{y,thin}/d_i=7.5$ and $5$ cases; the short reconnecting period (Fig.~2(a)) arises from the initial perturbation, but it tapers off quickly.

\begin{figure}
\includegraphics[width=8.5cm]{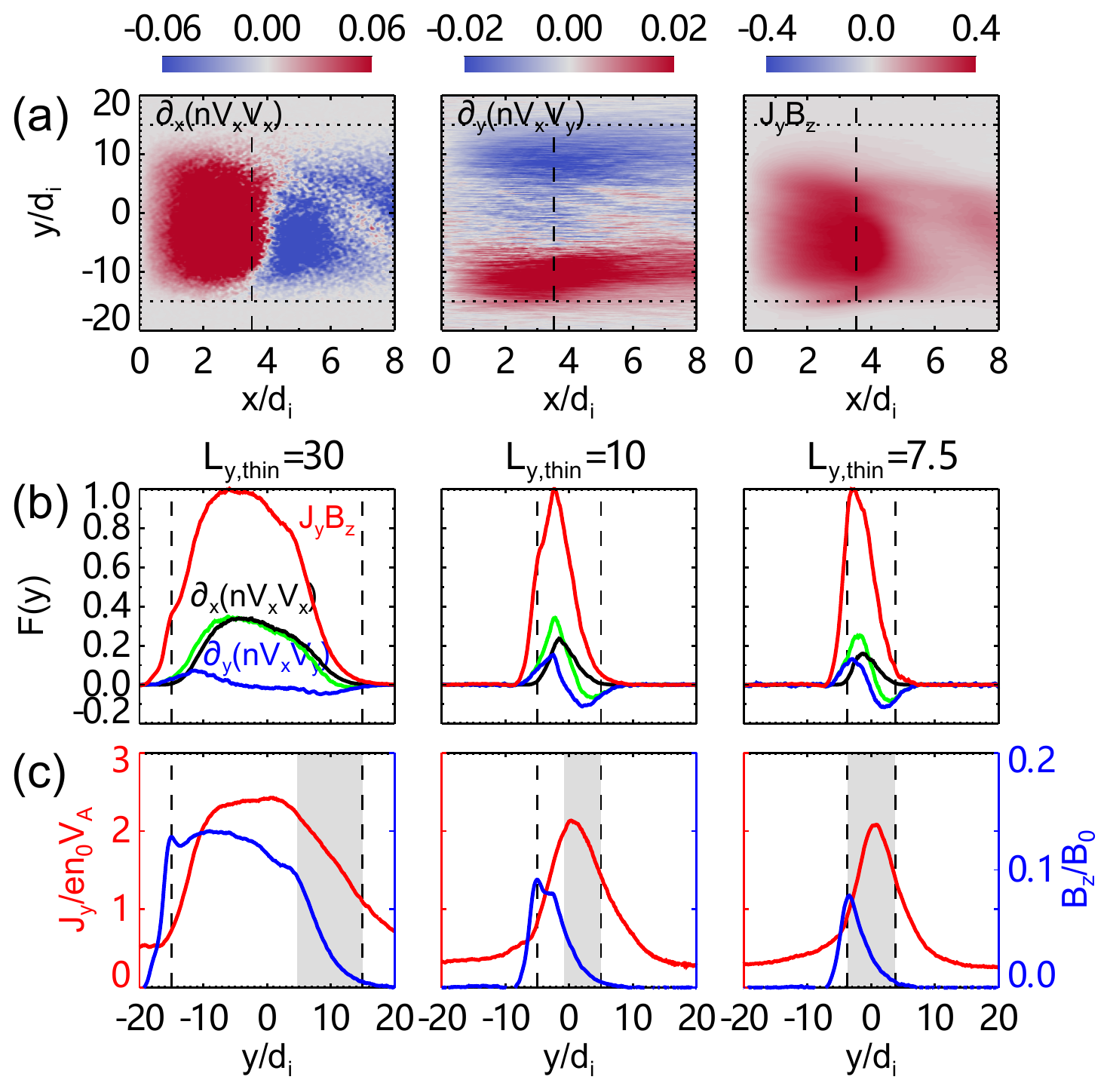} 
\caption {(a) The contour of the three terms in Eq.~(\ref{JxB}) in the $L_{y,thin}=30d_i$ case, the horizontal dashed lines bound the thin current sheet, the vertical dashed lines are $x=\xi$. (b) $\partial_x(nV_xV_x)$ in black, $\partial_y(nV_yV_x)$ in blue, $\partial_x(nV_xV_x)+\partial_y(nV_yV_x)$ in green and $J_yB_z$ in red averaged over $0<x<\xi$ in the cases with $L_{y,thin}/d_i=30$, 10, and 7.5. Quantities are normalized by the maximum of $J_yB_z$ in each case. (c) $J_y$ in red and $B_z$ in blue averaged over $0<x<\xi$. The gray shaded region indicates the suppression region. The dashed lines in (b) and (c) bound the thin current sheet region.}
\label{FG3}
\end{figure}

{\it Scaling of the outflow speed--}
To model the ion outflow speed, we analyze the momentum equation, $m_i[\partial_t(n{\bf V})+\nabla\cdot(n {\bf VV})]={\bf J}\times{\bf B}-\nabla\cdot {\bf P}$. Here $\bf V $ is the ion velocity, {\bf J} is the total current density, and $\bf P$ is the total pressure tensor. The electron inertia is neglected because of the smaller mass. The $J\times B$ force is the driver of reconnection outflow.  While gaining the bulk speed, downstream plasmas are also heated, developing a pressure gradient to decelerate outflows. Previous studies suggest that the ions gain thermal energy when they are picked up by the outflow magnetic fields \cite{drake09a}, resulting in an effective thermal velocity equal to the outflow speed; the temperature increase along the outflow can be modeled as $\Delta T_i\propto m_i V_{out}^2$. For simplicity, we will thus absorb the pressure gradient into the inertial term by assuming, $\nabla \cdot {\bf P} \propto m_i\nabla\cdot (n {\bf VV})$, that only makes a (constant) correction of order unity. In the steady state, the $\partial/\partial t$ term is negligible. These simplify the force balance in the $z=0$ plane into
\begin{equation}
\partial_x(nV_xV_x)+\partial_y(nV_yV_x) \propto J_y B_z.
\label{JxB}
\end{equation}

\begin{figure}
\includegraphics[width=7cm]{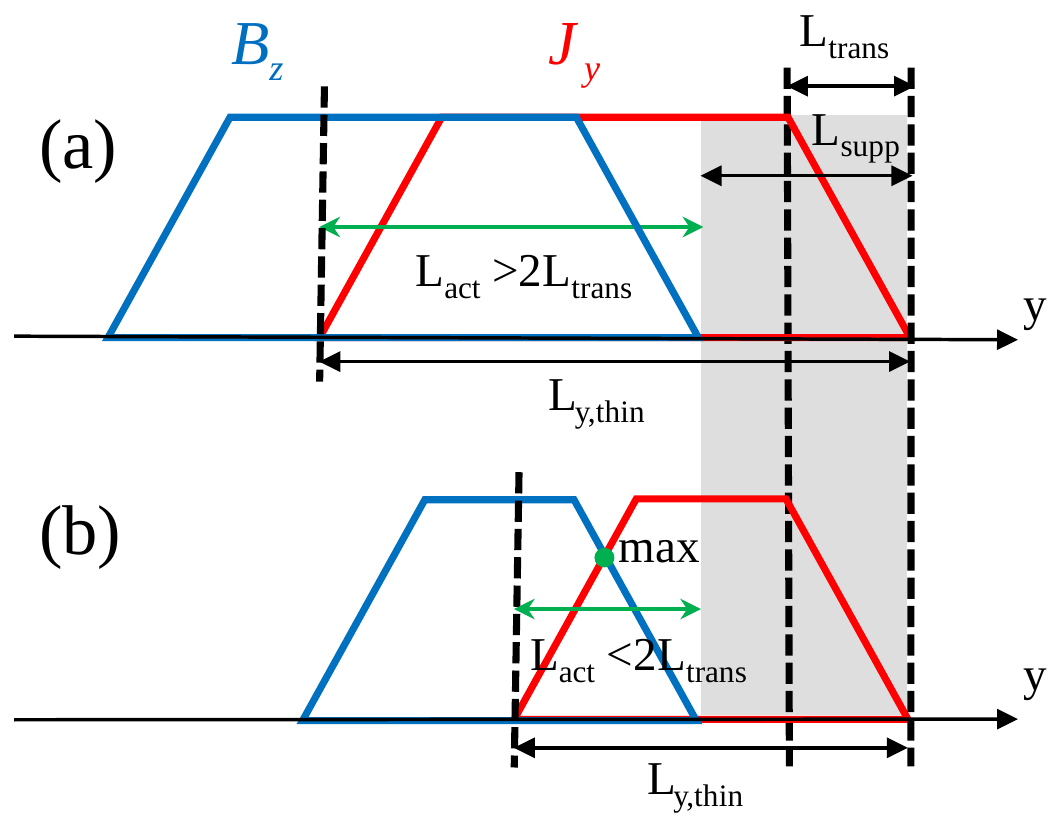} 
\caption {Modeling the reduction of $J_yB_z$ that results from the phase shift between $B_z$ and $J_y$. Colors are designed to match those in Fig.~3(c).}
\label{FG4}
\end{figure}

By balancing the 1st term of LHS to RHS, we recover the Alfv\'enic outflow speed \cite{parker57a}. The 2nd term of LHS has some effect at the edges, but it is the transport of $B_z$ that significantly reduces $J_yB_z$ on RHS, and thus the outflow speed in small $L_{y,thin}$ cases, as we will see. In Fig.~3(a), we plot the contour of the three terms in Eq.(\ref{JxB}) in the case of $L_{y,thin}=30d_i$. Note that the x-line is along the $x=0$ axis in Fig.~3(a). The average of these terms over $0<x<\xi$ are plotted in panel (b); here $x=\xi$ is the position (indicated by the vertical dashed line in panel (a)) where $V_{ix}$ reaches the maximum. The average is defined by $F(y)=\int_0^\xi f(x,y)dx/\xi$. In addition to the $L_{y,thin}=30d_i$ case, we also show cases with $L_{y,thin}= 10d_i$ and $7.5d_i$; they correspond to systems with a ``long active region'', ``short active region'' and ``no active region'' respectively. Because we concern the relation between the inertial term (green) and the $J\times B$ force (red), we normalize all the curves by the maximum of $J_yB_z$. In all three cases, the ratio between the inertial term and the $J\times B$ force remains a constant around 0.3, justifying the proportionality when absorbing $\nabla\cdot{\bf P}$. This additional factor of 0.3 gives us an outflow speed of $0.55 V_A$, which is still considered Alfv\'enic and is, in fact, often seen in 2D kinetic simulations. The dominant inertial term is $\partial_x(nV_xV_x)$, while $\partial_y(nV_yV_x)$ only becomes non-negligible \cite{arnold18a} at the edge of the outflow \footnote{If this term dominates, it should cause a suppression region on the dawn side instead.} in cases with a short $L_{y,thin}$.

For this reason, we just use $V_x^2\propto J_yB_z$ to model the peak ion outflow speed later. In the case without an active region, only weak $B_z$ could be generated, what observed in this case comes from a weak reconnection (that is not sustainable) driven by the initial perturbation.

The peak values of $J_yB_z$ in these three cases are $0.283$, $0.124$ and $0.083$, and this decrease is due to the transport of $B_z$ by electrons beneath the ion diffusion region (i.e., Hall effect). To illustrate this effect, we plot the averaged $J_y$ and $B_z$ in Fig.~3(c). $J_y$ correlates with the thickness of the currents sheet, which becomes larger in the electron diffusion region, but remains similar to the initial value in the outflow region. Meanwhile, the reconnected magnetic field $B_z$ at outflow is also similar in all cases (presumably because its value is limited by the upstream force balance \cite{yhliu17a}). Thus the peak values of $J_y$ ($B_z$) are considered similar in the three cases. 
 
However, while the peak of $J_y$ is located close to the center of the current sheet, the peak of $B_z$ is transported to the -y direction. This phase difference is more significant with a shorter $L_{y,thin}$, and this reduces the $J\times B$ force. The transport distance is basically the suppression region extent $L_{supp}$, and it is gray shaded in Fig.~3(c). Note that in the thick current sheet region (i.e., outside of the region between vertical dashed lines in (c)), $B_z$ decrease quickly because the electron drift speed is low in this region and $B_z$ will not be transported farther.

Motivated by these observations, we model the reduction of the $J\times B$ force in Fig.~4. For simplicity, we assume that both $B_z$ and $J_y$ have the same functional form $f(y)$. It has a plateau in the center and transition regions of scale $L_{trans}$ on both flanks. During reconnection, $J_y$ at the outflow region is centered within the thin current sheet, but $B_z$ is transported to the -y direction by distance $L_{supp}$. Thus $J_y=J_{y,max} f(y)$ and $B_z=B_{z,max} f(y+L_{supp})$. If $L_{act} \geq 2 L_{trans}$ as in Fig.~4(a), part of the plateau regions of $J_y$ and $B_z$ overlap, therefore $J_yB_z$ can reach the maximum value $J_{y,max}B_{z,max}$ and drives Alfv\'enic outflows with speed $\simeq V_A$; if $L_{act} < 2 L_{trans}$ as in panel (b), the plateau regions of $J_y$ and $B_z$ do not overlap and $J_y B_z$ will reach the maximum at where $f(y)=f(y+L_{supp})$. Using the linear interpolation, we obtain $(J_yB_z)_{max}=(L_{act}/2L_{trans})^2 J_{y,max}B_{z,max}$. Combined with $V_x^2\propto J_yB_z$ we model the reconnection outflow speed as
\begin{equation}
V_{x,max}= \left\{
\begin{array}{ll}
V_{m}\simeq V_A & \mbox{if $L_{act}\geq 2L_{trans}$} \\ 
(L_{act}/2 L_{trans}) V_m & \mbox{otherwise.}
\end{array} \right.
\end{equation}
where $L_{act}=L_{y,thin}-L_{supp}$. The maximum reconnection outflow speed scales linearly with the y-extent of the active region in the short x-line limit. When the active region is absent, the reconnection outflow should be suppressed.

\begin{figure}
\includegraphics[width=8.5cm]{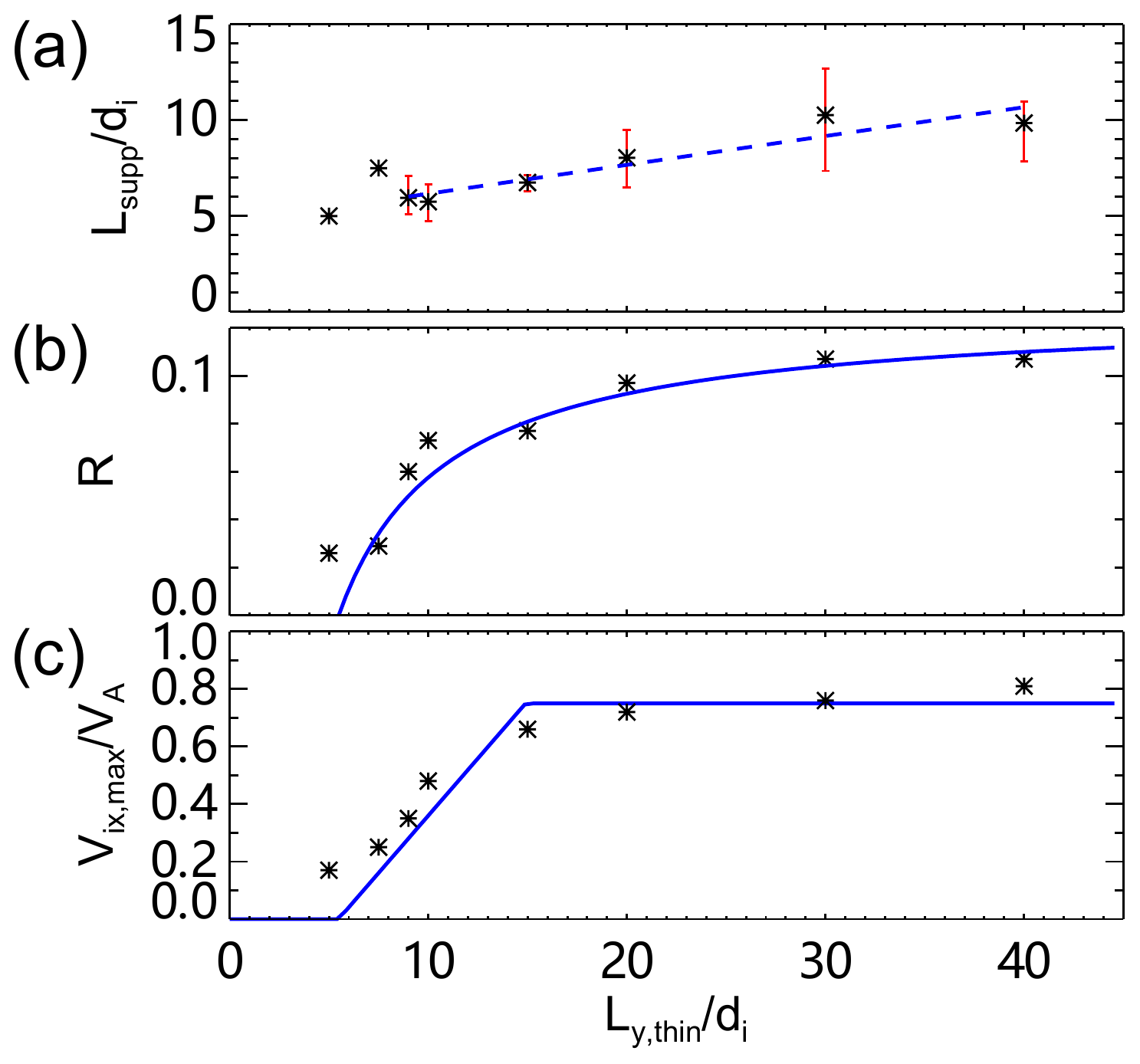} 
\caption {A comparison between the models (solid curves) and the simulation data (*). $L_{supp}$ in (a), $R$ in (b) and $V_{ix,max}$ in (c).}
\label{FG5}
\end{figure}

{\it Model-Data comparison and remarks--}
In this letter, we model the reconnection rate and outflow speed of 3D magnetic reconnection that has a limited x-line extent. 
We summarize the simulation data and compare them with our models in Fig.~5. Panel (a) shows the time-averaged suppression region extent in all cases. For $L_{y,thin}/d_i=7.5$ and $5$ cases, there is no active region so $L_{supp}=L_{y,thin}$. The red segments are the error bars resulting from the time variation of $L_{supp}$. In all cases, $L_{supp}$ appears to be on the order of $10d_i$ \cite{yhliu19a}. Fig.~5(b) plots the maximum reconnection rate $R$, and the solid curve is the prediction using Eq.(\ref{rate}), where $R^*$ is set to be $0.15V_A B_0$, and $L_{supp}$ uses the linear fit (dashed line) in Fig.~5(a). 

This curve traces the simulation results well. In panel (c), we plot the maximum ion outflow speed, and the solid curve is the prediction using Eq.(3), where we use $V_m=0.75V_A$, $L_{trans}=4d_i$ estimated from the simulation results,  $L_{act}=L_{y,thin}-L_{supp}$ with $L_{supp}$ again being the linear fit in 5(a). The model can explain most of the simulation result \footnote{Meyer et al. \cite{meyer15a} modeled the outflow reduction using 3D conservation laws. The reduction in their model comes from the different length scale between the inflow and outflow in the x-line direction. However, electron dynamics and Hall effect, which are essential in our study, were not considered in their model.}. The discrepancy is larger in the limit of small $L_{act}$ or $L_{y,thin}$ where the active region is absent; non-zero outflow velocities develop because of the initial perturbation. 
In general, $R^*\sim \mathcal{O}(0.1)$, $V_m\sim \mathcal{O}(V_A)$ and $L_{supp}\sim \mathcal{O}(10d_i)$ in our models, while $L_{trans}$ relates to the formation process of the thin current sheet; in Earth's magnetotail, possibilities include the length scale of ballooning instability  \cite{pritchett2011plasma, pritchett14a}, or localized streamers \cite{nishimura2016localized} that triggers reconnection. 

Other than the applications laid out in the introduction, magnetic reconnection is also considered as a primary atmospheric loss mechanism in planetary magnetotail [e.g.,\cite{zhang2012magnetic}]. Our model indicates that this mechanism might be invalid in planets with extremely small magnetotail, and the loss efficiency should take into account the suppression due to a small cross-tail scale. This new finding can also be important to the on-going ESA-JAXA mission, BepiColombo, that plans to map out the magnetic structure of Mercury’s magnetosphere.\\

\acknowledgments
Y. L. is grateful for support from NASA grant 80NSSC18K0754 and MMS mission. Simulations were performed at National Energy Research Scientific Computing Center (NERSC). 

\bibliography{paper}

\newpage

\end{document}